\documentstyle[12pt]{article}
\begin{document}

\begin{titlepage}
\begin{center}
{\Large {\bf Correlation Function in Ising Models \\}}
\vspace{2.0cm}
{\bf C.~Ruge$^a$, P. Zhu$^b$ and F.~Wagner$^a$ \\}
\vspace{1.5cm}
a) Institut f\"ur Theoretische Physik und Sternwarte\\
Univ. Kiel, D-24098 Kiel, Germany \\
E Mail: ruge@theo-physik.uni-kiel.de \\
\vspace{.5cm}
b) Dept. of Phys. Hangzou Univ.\\
Hangzou, 310028, P.R. China\\ 

\vspace{1.5cm}
\begin{abstract}
\noindent
We simulated the fourier transform of the correlation function of the 
Ising model in two and three dimensions using a single cluster algorithm with 
improved estimators. The simulations are in agreement with series expansion 
and the available exact results in $d=2$, which shows, that the 
cluster algorithm can succesfully be applied for correlations. We show as a 
further result that our data do not support a hypothesis of Fisher that in any 
$d=2$ 
lattice the fourier transform of the correlation function depends on the 
lattice generating function only. In $d=3$ our simulation are again in 
agreement with the results from the series expansion, except for the 
amplitudes $f_{\pm}$, where we find  $f_+/f_-=2.06(1)$.
\end{abstract}
\end{center}
\end{titlepage}

\section{Introduction}
\label{intro}
A simple way to characterize the behaviour of spin variables of an Ising model 
consists in the two point correlation function $g_{x,y}$.
Except for the trivial $d=1$ case a general expression has not been found. In 
$d=2$ are only the exact expressions for small and large separations known 
\cite{WU}. 
Apart from that most information comes from high temperature expansions 
\cite {FIS1,FIS3,DG3}. Universal quantities related to the critical 
behaviour with scaling dimension $0$ can be calculated from $\phi^4$ field 
theory \cite{MUEN,PRIV,BREZ}. 
Numerical simulations are difficult for several reasons. Direct simulation 
of $g_{x,y}$ requires an computational effort increasing with $N^2$, where 
$N$ is the number of lattice points. Within periodic boundary conditions $g$ 
is translational invariant and can be reduced to the fourier transform of 
$g_{0,x}$ .Since this still needs an effort increasing with $N$, only 
lattices of rather modest linear extension can be treated for $d>1$. For 
many applications the knowledge of the fourier transform along one or two 
directions is sufficient which renders us a feasible problem. Even within 
that restriction success depends on the algorithm. In general one may adopt 
a single spin update (Metropolis or Heatbath algorithm \cite{MET}) or a 
cluster 
algorithm \cite{SW,UW}. For thermal quantities the smaller autocorrelation 
time of the 
latter is compensated by the better vectorization of the former algorithm 
\cite{ITO}. For the fourier transform of the correlation function one has 
to measure 
$|\sum_x e^{ikx}\sigma_x|^2$. Since both the spin variable $\sigma_x$ and 
the phase factor vary rapidly with $x$, the resulting fluctuations prevent 
any accurate measurement.  The problem of   
fluctuation is particular striking approaching the critical temperature. 
The improved estimators in a cluster algorithm  
have much less variation and lead to more accurate results. The 
aim of the present paper is to show that the correlation function can be 
reliably determined with the single cluster algorithm of Wolff \cite{UW} by 
comparing our results with known exact results in $d=2$ and the results from 
series expansion in $d=2$ and $d=3$ \cite{FIS1,FIS3,LIU}.\\
The paper is organized in the following way. In Section \ref{corr} we 
collect the necessary formalism on the correlation function. The cluster 
algorithm is described in section \ref{clus} and section \ref{result} 
contains 
the results of our simulations. In section \ref{conc} we give our 
conclusions.


\section{Correlation functions}
\label{corr}
The connected pair correlation function $g^c_{x,y}$ for an Ising model is 
defined as 
\begin{equation}\label{conn}
g^c_{x,y} = \left\langle \sigma_x \sigma_y \right \rangle - m^2 
\Theta(T_c-T)
\end{equation}
$x,y$ denote the sites of a d-dimensional simple cubic lattice of linear 
dimension $L$ with periodic boundary conditions and $\left\langle A \right 
\rangle$ means the thermal average at temperature $T$ taken with an energy 
$E$
\begin{equation}
E= \frac{1}{2}\sum_{x,y} \delta_{|x-y|,1} \sigma_x \sigma_y
\end{equation}
where $\delta_{|x|,1}$ projects on next neighbours of $x=0$.\\
Below the critical temperature a magnetization term has to be subtracted 
from $\left\langle \sigma_x \sigma_y \right \rangle$ in order to obtain a 
reasonable behaviour at large distances $|x-y|$.  $g^c_{x,y}$ can be 
considered as a $L^d \cdot L^d$ matrix. Usually its inverse (the proper 
vertex function in field theory) has a simpler behaviour. By means of 
translation invariance we can write 
\begin{equation}\label{FOU}
\left(g^c\right)^{-1}_{x,y} = \frac{1}{L^d}\sum_ke^{ik\left(x-y 
\right)}\frac{1}{\hat g(k)}
\end{equation}
$k$ is a $d$ dimensional vector whose components $k_i$ are restricted to the 
first Billouin zone $|k_i|\le \pi$ and have the form $\frac{2 \pi 
}{L} n_i$ with integer $n_i$. The positive function $\hat{g} (k)$ is simply 
the fourier transform of  $g^c_{0,x}$. For $|k|<<1$ it has the expansion
\begin{equation}
\label{effective}
\frac{1}{\hat{g}(k)} = \frac{1}{\chi (T)}\left(1+\xi^2k^2+ ...\right)
\end{equation}
where $\chi(T)$ denotes the susceptibility in units of the Curie 
susceptibility and $\xi(T)$ the effective correlation length.
Near $T \sim T_c$ we expect the following dependance on the scaling variable 
$\tau = 1 - \frac{T}{T_c}$:
\begin{eqnarray}\label{indices}
\chi(T) &=& C_{\pm}|\tau|^{-\gamma}\nonumber\\
\xi(T)  &=& f_{\pm}|\tau|^{-\nu}\nonumber\\
m       &=& B \tau^{\beta}.
\end{eqnarray}
The $\pm$ signs in the amplitudes refer to $T 
\stackrel{\scriptstyle >}{\scriptstyle <}T_c$. In $d=2$ the critical 
indices and the amplitudes $C$ and $B$ are known exactly 
\cite{YANG,ONS,BAR,ABR,BAX}, but the exact 
known $f^t_{\pm}$ refer 
to the true correlation length  $\xi^t$ which is defined by the exponential 
decay of 
the correlation function $\ln g_{x,y}\sim -|x-y|/ \xi^t$ for large 
separations and may 
differ especially for $T<T_c$ from  
$\xi$. Theoretical information on $f_{\pm}$ in $d=2$ and $d=3$ comes from 
series expansion 
\cite{FIS1,FIS3} and on the ratio $f_+/f_-$ in $d=3$ from renormalized 
perturbation theory 
\cite{MUEN}.
 As discussed in the 
next section $\hat{g}(k)$ can be estimated by numerical simulations only at 
special values of the arguments. In the following we adopt the choices 
$k_R=k_0(1,1,..1)$ (radial direction) and $k_L=k_0(1,0,..0)$ (linear 
direction). This corresponds to two functions $g_L(k_0)$ and  $g_R(k_0)$ 
depending only on a scalar argument $k_0$:
\begin{eqnarray}
\label{gLR}
g_L(k_0) &=& \hat{g}(k_L)\nonumber\\
g_R(k_0) &=& \hat{g}(k_R)
\end{eqnarray}
Due to the rotationally invariant form of eq(\ref{effective}) $\chi$ and $\xi$ 
can be determined from one of the functions (\ref{gLR}). In $d=3$ we will 
restrict our numerical analysis to the small $k$ region.
In $d=2$ we can test in addition the following hypothesis stated by Fisher 
et. al. \cite{FIS1}. Motivated by the results of the high 
temperature expansion they 
conjectured, that for all $d=2$ Ising models $\hat{g}(k)$ is a function of 
the lattice generating function $q(k)$ only which is given on a square 
lattice by
\begin{equation}
q\left(k\right) = 2 \left( cos\left(k_1 \right) + 
cos\left(k_2\right)\right).
\end{equation}
Since $\hat{g}(q(k))$ is a function of one variable only , $g_L$ and $g_R$ 
must satisfy the following relation
\begin{equation}\label{RL}
g_R\left(k'_0 \right) = g_L\left(k_0 \right)
\end{equation}
provided $k'_0$ is connected to $k_0$ by
\begin{equation} 
sin^2\left(k_0/2 \right) = 2 sin^2\left(k'_0/2 \right).
\end{equation}
Even if eq(\ref{RL}) would not be true, expansion of $1/\hat{g}$ in powers of 
$q$ consists in a very accurate representation of $g^c_{x,y}$. Writing
\begin{equation} \label{exp}
\frac{1}{\hat{g}(k)} = 1 - \sum_n H_n\left( T \right) \left(q \left(k\right) 
\right)^n
\end{equation}
only $H_n$ with $n\le 2$ are present up to $o(1/T)^{12}$. Converting 
eq.(\ref{exp}) in position space we get from eq.(\ref{FOU})
\begin{eqnarray} \label{gPOS}
\left(g^c\right)^{-1}_{x,y}&=& a_0\left(T\right)\delta_{x,y} - 
 a_1\left(T\right)\delta_{|x-y|,1}\nonumber\\
 &-&
 a_2\left(T\right)\left(\delta_{|x-y|,2}+ 2 \delta_{|x-y|,\sqrt{2}}\right) +
o\left(1/T\right)^{12}
\end{eqnarray}
where $\delta_{|x-y|,r}$ projects on points $x,y$ having euclidean 
distance $r$. 
The high temperature expansion of ref \cite{FIS1} for $a_i(T)$ are given 
for convenience 
in appendix A. $a_2(T)$ is of order $(\frac{1}{T})^{10}$ and therefore 
very small. If the expansion (\ref{gPOS}) is valid, the coefficients $a_i$ can 
be determined from low order moments of the function (\ref{gLR})
\begin{eqnarray} 
a_0\left(T\right)&=& \frac{1}{2L} \sum_{k_0} \left[ \frac{2+cos 
k_0}{g_R\left(k_0\right)} + \frac{cos 
k_0}{g_L\left(k_0\right)}\right]\nonumber\\
a_1\left(T\right)&=& \frac{1}{2L} \sum_{k_0}  \frac{cos
k_0}{g_R\left(k_0\right)}\nonumber \\
a_2\left(T\right)&=& \frac{1}{L} \sum_{k_0}  \frac{cos        
2k_0}{g_L\left(k_0\right)} \label{a(T)}.
\end{eqnarray}
Note, that the sum $\sum_{k_0}$ is only one dimensional.
For $1/T<0.4$ the first two terms in eq.(\ref{gPOS}) give a very accurate 
representation of $\left(g^c\right)^{-1}_{x,y}$, which may be useful in 
applications \cite{SCH}.


\section{Numerical methods}
\label{clus}
A direct simulation of the disconnected correlation function
\begin{equation}\label{gSIM}
g_{x,y} = \left\langle \sigma_x \sigma_y \right \rangle
\end{equation}
is prohibitive both from storage requirement and computational effort, 
even the latter can be reduced by Boolean operations (since 
$\sigma_x$ are two valued quantities) and multi spin update \cite{BHAN}.
Taking translation invariance within periodic boundary conditions into 
account only $g_{0,y}$ is needed, or
\begin{equation}\label{g0y}
g_{0,y} = \frac{1}{L^d} \sum_{x'} \left\langle \sigma_{x'}
 \sigma_{y+x'} \right \rangle  
\end{equation}
Having cured the storage problem the computational problem remains that 
eq.(\ref{g0y}) requires $O(L^d)$algebraic operations per single spin update. 
For 
the observables discussed in section \ref{corr} not the full fourier 
transform of $g$ is needed but rather ${g}_L(k_0)$ or ${g}_R(k_0)$ 
with $k_0\not= 0$
\begin{equation}\label{gFOU}
{g}_{R,L}\left(k_0\right) = \frac{1}{L^d} \left\langle|\sum_x
e^{-i\left(x\cdot k_{R,L}\right)}\sigma_x|^2\right\rangle.
\end{equation}
The value at $k_0=0$ has to be determined by an extrapolation $k_0 \to 0$. The 
difference of this extrapolated value and the actual value of the r.h.s. of 
eq.(\ref{gFOU}) gives $m^2L^d$. Since $k_0$ in eq.(\ref{gFOU}) can assume 
only $L$ different values the computational effort can be tolerated. 
However, eq.(\ref{gFOU}) will still not lead to any meaningful result if one 
uses a conventional MC Method with single spin update. The reason are the 
cancellations in the sum involved in eq.(\ref{gFOU}) necessary to obtain a 
finite 
value for ${g}_{R,L}$. The corresponding fluctuations in each 
single measurement prevent any convergence of the average value inside 
present computing facilities. These fluctuations can be avoided by using a 
cluster algorithm. Its merit is not so much the reduced autocorrelation time 
between different configurations because this is balanced by the better  
vectorization of the local algorithms \cite{ITO}, but the existence of 
improved 
estimators for spin expectation values as (\ref{gSIM}). For the single cluster 
method the estimator for the correlation function can be written as (see 
Appendix B)
\begin{equation}\label{gIMP}
{g}_{R,L}\left(k_0\right) = \left\{ \frac{1}{s} |\sum_{x \in C}
e^{-i\left(x\cdot k_{R,L}\right)}|^2 \right\}.
\end{equation} 
\{ \} means an average over all single clusters $C$ and $s$ denotes the number 
of sites in $C$. Comparing 
eq.(\ref{gFOU}) and eq.(\ref{gIMP}) we notice the absence of the spin factor $\sigma_x$ 
which is due to the fact that all cluster spins have the same value. Moreover 
the sum over $x$ extends only over the cluster instead of the whole lattice 
as in eq.(\ref{gFOU}). Since the size of a cluster in the high temperature 
phase 
is governed by the correlation length $\xi$, the phase factor will not vary 
substantialy for $k_0\xi$ of $O(1)$ which is precisely the region we are 
interested in an application of eq.(\ref{gIMP}). Since the cancellations are 
only 
needed to render ${g}_{R,L}$ to be an decreasing function of $k_0$ they are no 
problem at all. Note that this improvements hold only for cases where the 
magnetization is small. For $T<T_c$ and finite $m$ a fraction $O(m)$ of the 
clusters will extend over the whole lattice and the same problems as in the 
single spin MC will reappear. Therefore we can use our method in the ordered 
phase in a narrow 
region of $T_c-T$ only.\\
For an error  estimate one has to keep in mind that the autocorrelation 
time $t_a$ for a cluster algorithm even much smaller than for single spin 
update may not be small absolutely. Therefore we divide the number of 
generated clusters into $n_B$ blocks of size $n_s$. In each block the average 
of any observable is taken. The final mean value of the observable and its 
error are calculated from the different means of the blocks. This procedure 
will protect against autocorrelation times $t_a<n_s$. Typically we take a 
ratio $n_s/n_B \sim 100 - 500$. Also functions of observables 
(f.e.$g^{-1}_L$) including their error may be easily estimated.

\section{Results from numerical simulations}
\label{result}
In $d=2$ exact expressions for the behaviour of $\hat{g}^{-1}(k)$ at small 
values of k are known. For a comparism we used the cluster algorithm on a
$160\times160$ lattice to simulate  $1/g_{R,L}$. Typically we generated 
$10^5$ clusters in the high temperature phase and $10^4$ clusters in the 
ordered phase. 
The range of temperatures $\tau = \pm 0.1$, $\tau = \pm 0.08$, $\tau = \pm 
0.05$ and  $\tau = \pm 0.03$ was motivated because the correlation length must be 
small compared to the lattice size in order to avoid any finite size 
corrections. In Fig. 1 we show   ${g}^{-1}_R(k)\tau^{-7/4}$ as 
function of $sin^2(k/2)$ for various temperatures $T>T_c$. Superimposed are 
straight line fits to points restricted to $\xi^2k^2<1$. For $\tau\ge0.08$ 
the data exhibit a linear behaviour in $sin^2k/2$ in the whole $k$ region. 
Approaching $T_c$ deviations 
from this behaviour become important in  
agreement with the high temperature expansion. ${g}^{-1}_L(k)$ 
exhibits the same behaviour. Since the statistical accuracy of $g_R$ is much 
better and the extrapolation of $g_L$ to $k \to 0$ is compatible  with the one 
obtained from $g_R$ we discuss only the latter. The straight line fits yield 
 us values for $\chi$ and $\xi^2$ 
according eq.(\ref{effective}). These fits are represented by the lines in Fig. 
1. The resulting $\chi$ and $\xi$ are shown in Fig. 2 on a double log scale 
as function of $\tau$. The lines give the exact theoretical result 
\cite{BAR,ABR,BAX}, which 
agree well with the extrapolations of our data. The same analysis can be 
done for $T<T_c$. Due to the non vanishing magnetization less 
statistic is available and the quality is poorer as compared to $T>T_c$. 
 The deviation from a straight line behaviour sets in much earlier than for 
$T>T_c$, as can be seen from the big difference of data and the 
corresponding lines fitted to the small $k$ region. From the fits we obtain 
$\chi$ and $\xi$ which are also shown in Fig.2. 
Again we notice the agreement with the exact results \cite{BAR,ABR,BAX}. 
The value of 
$g^{-1}_R(0)$ can be translated via eq.(\ref{conn}) in a value for the 
magnetization. The values for $m$ are divided by the theoretical value 
\cite{YANG,ONS} 
\begin{equation}\label{magth}
m_{th} = \left( 
\frac{sinh^4\left(2/T\right)-1}{sinh^4\left(2/T\right)}\right)^{1/8}.
\end{equation}
As one sees from Fig.~2, this ratio is compatible with 1 inside the small 
errors. In this case straight line fits to m are not very meaningful, since 
eq.(\ref{magth}) predicts a substantical deviation from the power law 
(\ref{indices}) 
in our range 
of $\tau$ values. To demonstrate the statistical accuracy we fit our values 
of $\chi$ and $\xi$ with the power laws (\ref{indices}). In fit A we fix the 
exponents to the theoretical values $\gamma = 7/4$ and $\nu = 1$, in fit C 
we fit in addition the exponents $\gamma$ resp. $\nu$ simultaneously to the 
data for $T\stackrel{\scriptstyle >}{\scriptstyle <}T_c$. The resulting 
parameters are displayed in table 1 
together with the known exact values, resp. the values from the series 
expansion for the effective correlation length $f_{\pm}$ from ref 
\cite{FIS1,FIS3}. Both 
agree within the statistics which shows that the cluster algorithm 
leads to reasonable values for the correlation function.\\
In $d=2$ we can test apart from the behaviour near $T_c$ also the results of 
the high temperature expansion. If the coefficients $H_n$ for $n>2$ in the 
expansion of $\hat{g}^{-1}$ can be neglected , $a_0$ and $a_1$ can be 
determined from $g_L$ and $g_R$. The experimental value of coefficients  
$a_0$ and $a_1$ are obtained by eq.(\ref{a(T)}). These values divided by the 
series expansions  (\ref{APP}) are shown in Fig. 4 as function of $1/T$ for 
various values of the lattice size $L$. Below $1/T = 0.40$ the first two 
coefficients in the expansion (\ref{gPOS}) computed by the high temperature 
series give a good description of the correlation function which can be used 
in practical applications \cite{SCH}. The deviations above $0.4$ are not due 
to 
finite size effects. They signal not so much a breakdown of the series 
expansion for $a_{0,1}$ but rather the appearance of higher coefficients 
$a_n$ which spoil the relations (\ref{a(T)}) between $g^{-1}$ and $a_{0,1}$. 
This is corroborated by the values of $a_2$ obtained by eq.(\ref{a(T)}) 
which are 
zero below $1/T = 0.40$ and quickly exceed the value expected from 
eq.(\ref{APP}) by an 
order of magnitude. Whereas finite size corrections are negligiable for 
leading effects discussed up to now, a test of the Fisher hypothesis 
$\hat{g}(k)$ being a function of the lattice generating function only is 
very sensitive on small finite size corrections. This is because any 
deviation from the relation (\ref{a(T)}) can be a small effect only and the 
cancellations necessary to suppress the low order $1/T$ terms in $H_n$ are 
no longer effective at order $1/L$. We can control these corrections by 
the number of clusters with loop number \cite{RDW} non zero (clusters wrapping 
around the lattice at least once). Requiring the fraction of those clusters 
to be less than $10^{-6}$ we have to choose on a $160 \times 160$ lattice an 
inverse temperature not larger than $0.42$. In order to make a small effect 
visible, we divide both $g^{-1}_L(sin^2 \frac{k_0}{2})$ and 
$g^{-1}_R(\frac{1}{2}sin^2\frac{k'_0}{2})$ by 
a common linear factor R
\begin{equation}
R = 0.005 + 2.63 sin^2k_0/2
\end{equation}
obtained by a fit to $g^{-1}_L$. In Fig.~5 
$g^{-1}_L(sin^2\frac{k_0}{2})/R$ and 
$g^{-1}_R(\frac{1}{2}sin^2\frac{k'_0}{2})/R$ are shown as function of 
$sin^2\frac{k_0}{2}$. If $g^{-1}$ 
would be linear in $sin^2\frac{k_0}{2}$, both data points would fall 
on a constant 
line at 1. The deviation from 1 signalizes the appearance of higher 
coefficients $H_n$ with $n\ge 2$ as expected from the series expansion. If 
the hypothesis (\ref{a(T)}) is true, both sets of data points have to agree, 
which is obviously not the case inside our errors. Since at the level of 
accuracy of $10^{-3}$ finite size corrections can be neglected, our data are 
in disagreement with Fishers conjecture.\\
For $d=3$ we restrict ourselves to the behaviour near $T_c=1/0.22165$ 
\cite{FER,BAI}.
The computational effort increases rapidly with $L$, especially in the 
ordered phase at $T<Tc$. The data for a $40^3$ lattice shown in Fig. 6 
needed $\sim 30$h on a CRAY YMP. Since the correlation lengths are smaller 
in $d=3$, we can cover the same range of $\tau$ as in $d=2$. As before the 
function  
$g_R(k_0)= \hat{g}(k_0,k_0,k_0)$ has smaller errors than $g_L$. We do not 
 show a plot of $g_R$, since its behaviour is very similar to the 
$d=2$ case. Performing linear fits in the small $k$ region we obtain 
$\chi(\tau)$,$\xi(\tau)$ and $m(\tau)$ shown in Fig. 6 on a double log 
scale. From the linear behaviour we see that the power laws predicted by 
eq.(\ref{indices}) are well satisfied. To obtain values for exponents and 
amplitudes we performed the following fits with or without correction to 
scaling:
\begin{eqnarray}
\chi_{\pm} &=& C_{\pm}|\tau|^{-\gamma} \left(1+a_{\pm}|\tau| \right) \\
\xi_{\pm} &=& f_{\pm}|\tau|^{-\nu} \left(1+a'_{\pm}|\tau| \right) \\
m &=& B \left(-\tau\right)^{\beta} \left(1+b_{\pm}|\tau| \right) .
\end{eqnarray}
In principle the correction terms should be parametrized by nonlinear power 
laws, but within our statistics a determination of an exponent in the 
nonleading term is not possible. We checked 
that replacing $|\tau|$ by $|\tau|^{1/2}$ in the correction does not change 
the values of exponents or amplitudes. In fit A we fix the exponents to the 
generally accepted values \cite{FER,BAI,LIU} and 
leave out the corrections which are allowed in fit B. The fits C and D are 
similar to A and B except that also the exponents $\gamma$,$\nu$ and $\beta$ 
are free parameters. The resulting values and the corresponding  
$\chi^2/DoF$ are given in table II. In the fits to $\chi$ and $m$ correction 
terms are definiteley needed, whereas for $\xi$ inclusion of the latter does 
not improve $\chi^2$. In the case of the susceptibility variation of 
$\gamma$ does not reduce $\chi^2/DoF$. Therefore we believe that fit B is the 
most reliable 
one and is also shown in Fig. 6. The values for the amplitudes are in 
agreement with the series expansion (last column in table II). Variation of 
the exponent $\nu$ for the correlation length leads to a value very close to 
the accepted value of 
$\nu$. 
Since including corrections in fit B did neither change the values of 
$f_{\pm}$ nor improve $\chi^2/DoF$ fit D should not be trusted so much. In 
this case we adopt fit C as the most reliable one. A common feature of fit 
A,B and C is a consistent $5\%$ deviation in $f_-$ from the series 
expansion. The universal ratio (from fit C) $f_+/f_-$
\begin{equation}
f_+/f_- = 2.063(12)
\end{equation}
disagrees with ref \cite{LIU,BREZ}, but is in agreement with a recent 
calculation using 
renormalized perturbation expansion in $d=3$ \cite{MUEN} and with 
experimental data \cite{PRIV}. In the case of $m$ fit D is 
definitely preferred due to the acchieved $\chi^2/DoF$. Both exponent 
$\beta$ and the amplitude $B$ are somewhat smaller as compared to the series 
expansion but are consistent within the errors. In table II our final values 
for the amplitudes are underlined. With the exception of $f_-$ the cluster 
Monte Carlo method yielded values compatible with the series expansion with 
similar accuaracy.
\section{Conclusion}
\label{conc}
Due to fluctuations the correlation function $g$ in Ising models for $d>1$ 
cannot be determined with local Monte Carlo methods except for rather modest 
lattice sizes. Since the improved estimators in a cluster algorithm exhibit 
much less fluctuation the fourier transform of $g$ can be measured. We used 
the known results in $d=2$ to demonstrate the validity of our method. As a 
byproduct we show that a hypothesis of Fisher may not be true. In $d=3$ our 
simulations agree with the values expected from series expansion except for 
the amplitude $f_-$ of the effective correlation length in the ordered 
phase. The universal ratio $f_+/f_-=2.06(1)$ is in agreement with recent 
field theoretical estimates.\\
\\
\\
{\Large{\bf Acknowledgements:\\}}
\\
One of us (P.Z.) thank the University of Kiel for a grant. 
Computing time on a CRAY YMP was provided by HLRZ at J\"{u}lich.

\begin{appendix}
\section{High temperature expansion}
Taking the fourier transform of eq.(\ref{gPOS}) and comparing the coefficients of 
$cos(k_1n)$, $cos(k_2m)$ with eq.(\ref{exp}) one finds
\begin{eqnarray}
a_0 &=& 1- H_0(T)- 4 H_2(T)\nonumber \\
a_1 &=& H_1(T) \nonumber  \\
a_2 &=& H_2(T).
\end{eqnarray}
$H_i$ have been given in \cite{FIS1} in terms of powers of $v=tgh(1/T)$ 
which leads to
\begin{eqnarray}
a_0 &=& 1 + 4v^2 + 12 v^4 + 44 v^6 + 188v^8 + 836v^{10} + o(v^{12})\nonumber \\
a_1 &=& v(1 + v^2 + 5v^4 + 21v^6 + 96v^8 + 401v^{10}) + o(v^{13}) \nonumber \\
a_2 &=& 4v^{10} + o(v^{12})\label{APP}
\end{eqnarray}   

\section{The improved estimator}
The fourier transform of eq.(\ref{g0y}) is
\begin{equation}
\hat{g}(k) = \frac{1}{L^d}\sum_{x,y}e^{ik(x-y)}\left \langle \sigma_x \sigma_y 
\right \rangle
\end{equation}
For a Swendsen Wang \cite{SW} cluster decomposition of the lattice we have
\begin{equation}
\makebox[2.5cm][r]{$\left \langle \sigma_x \sigma_y \right \rangle$} = 
\sum_{C,C'}\Delta_x(C) 
\Delta_y(C') \left \langle \sigma_C \sigma_{C'} \right \rangle
\end{equation}
$\Delta_x(C)$ takes the value $1$ if $x \in C$ and $0$ otherwise.
Noting that cluster spins are independent variables
\begin{equation}
\makebox[.8cm][r]{}=\sum_C\Delta_x(C) \Delta_y(C)
\end{equation}
so
\begin{equation}
\makebox[2.cm][r]{$\hat{g}(k)$} = \frac{1}{L^d}\left\{\sum_C |\sum_{x \in C} 
e^{ikx}|^2 
\right\}_{SW}.
\end{equation}
The $1/L^d\sum_C$ in the Swendsen Wang algorithm translates into $1/s$ 
for a single cluster algorithm \cite{RDWW} 
so
\begin{equation}
\hat{g}(k) = \left \{ \frac{1}{s}|\sum_{x \in C }e^{ik\cdot x}|^2 \right \}
\end{equation}
from which we obtain eq.(\ref{gIMP}).
\end{appendix}

\newpage
\begin{center}
{\bf table I\\}
\vspace{.4cm}
\begin{tabular}{|l|c|c|c|}
\hline
&A&C&\\
\hline
\hline
$\gamma$&7/4&1.753(4)&7/4\\ \hline
$C_+$&0.98(2)&0.97(9)&0.962582\\ \hline 
$C_-$&0.0247(3)&0.025(3)&0.025537\\ \hline
\hline
$\nu$&1&0.99(3)&1\\ \hline
$f_+$&0.558(4)&0.55(2)&$0.56702(5)^{\cite{FIS1}}$\\ \hline
$f_-$&0.16(3)&0.18(2)&$0.175(5)^{\cite{FIS3}}$\\ \hline
\hline
\end{tabular}
\end{center} 
\vspace{1.5cm}

\begin{center}
{\bf table II\\}
\vspace{.4cm}

\begin{tabular}{|l|c|c|c|c|c|}
\hline
&A&B&C&D&series$^{\cite{LIU}}$\\
\hline
\hline
$\gamma$&1.237&\underline{1.237}&1.168(10)&1.235(44)&1.237\\ \hline
$C_+$&1.198(5)&\underline{1.093(13)}&1.432(33)&1.12(19)&1.103(1)\\ \hline
$C_-$&0.191(4)&\underline{0.211(11)}&0.237(11)&0.219(41)&0.223(3)\\ \hline
$\chi^2/DoF$&81/18&\underline{18/16}&29/17&18/15&-\\ \hline
\hline
$\nu$&0.629&0.629&\underline{0.628(1)}&0.618(6)&0.629\\ \hline
$f_+$&0.4997(2)&0.4995(8)&\underline{0.501(2)}&0.522(11)&0.496(4)\\ \hline
$f_-$&0.2415(8)&0.238(2)&\underline{0.243(1)}&0.251(8)&0.251(1)\\ \hline
$\chi^2/DoF$&64/16&62/14&\underline{63/15}&59/13&-\\ \hline
\hline
$\beta$&0.330&0.330&0.301(2)&\underline{0.319(5)}&0.330\\ \hline
$B$&1.632(2)&1.686(4)&1.478(9)&\underline{1.608(37)}&1.71(2)\\ \hline
$\chi^2/DoF$&263/11&35/9&35/10&\underline{24/8}&-\\ \hline
\end{tabular}
\end{center}

\newpage
{\bf{\large Figure captions\\}}
\\
{\bf Fig.~1:} The inverse fourier transform of the correlation function 
$g_L^{-1}(k)$ above $T_c$ as function of $sin^2\frac{k}{2}$ for $\tau 
=0.1,0.08,0.05$ and $0.03$ (from bottom to top). The straight lines are fits 
to the data for $\xi^2k^2<1$.\\
\\
{\bf Fig.~2:} $d=2$: The susceptibility $\chi$, the effective correlation 
length 
$\xi$ and the magnetization $m$ divided by eq.(\ref{magth}) as function of 
$|\tau|$ on a double log scale. $\pm$ indicates the sign of $T-T_c$. Solid 
lines represent the fit C (fit A) to $\xi$ ($\chi$) described in the text. 
The dotted line is the constant 1.\\
\\
{\bf Fig.~3:} The inverse fourier transform of the correlation function
$g_L^{-1}(k)$ below $T_c$ as function of $sin^2\frac{k}{2}$ for $\tau
=-0.1,-0.08,-0.05$ and $-0.03$ (from bottom to top). The straight lines 
are fits
to the first two $k$ values with $k\not=0$.\\
\\
{\bf Fig.~4:} The coefficients $a_0$ and $a_1$ from eq.(\ref{a(T)}) divided by 
the series expansion (\ref{APP}) for different values of the lattice size $L$ 
as function of $1/T$. The critical temperature is indicated by the arrow.\\
\\
{\bf Fig.~5:} The linear $g_L(k_0)$ and the radial $g_R(k'_0)$ correlation 
function at $1/T=0.42$ as function of $sin^2\frac{k_0}{2}$. If Fishers' 
hypothesis holds both have to be equal.\\
\\
{\bf Fig.~6:} $d=3$: The susceptibility $\chi$, the effective correlation 
length 
$\xi$ and the magnetisation $m$ as function of $|\tau|$ on a double log 
scale. $\pm$ indicate the sign of $T-T_c$. The straight lines correspond to 
fit B (fit C, fit D) for $\chi$ ($\xi$,$m$).\\
\\








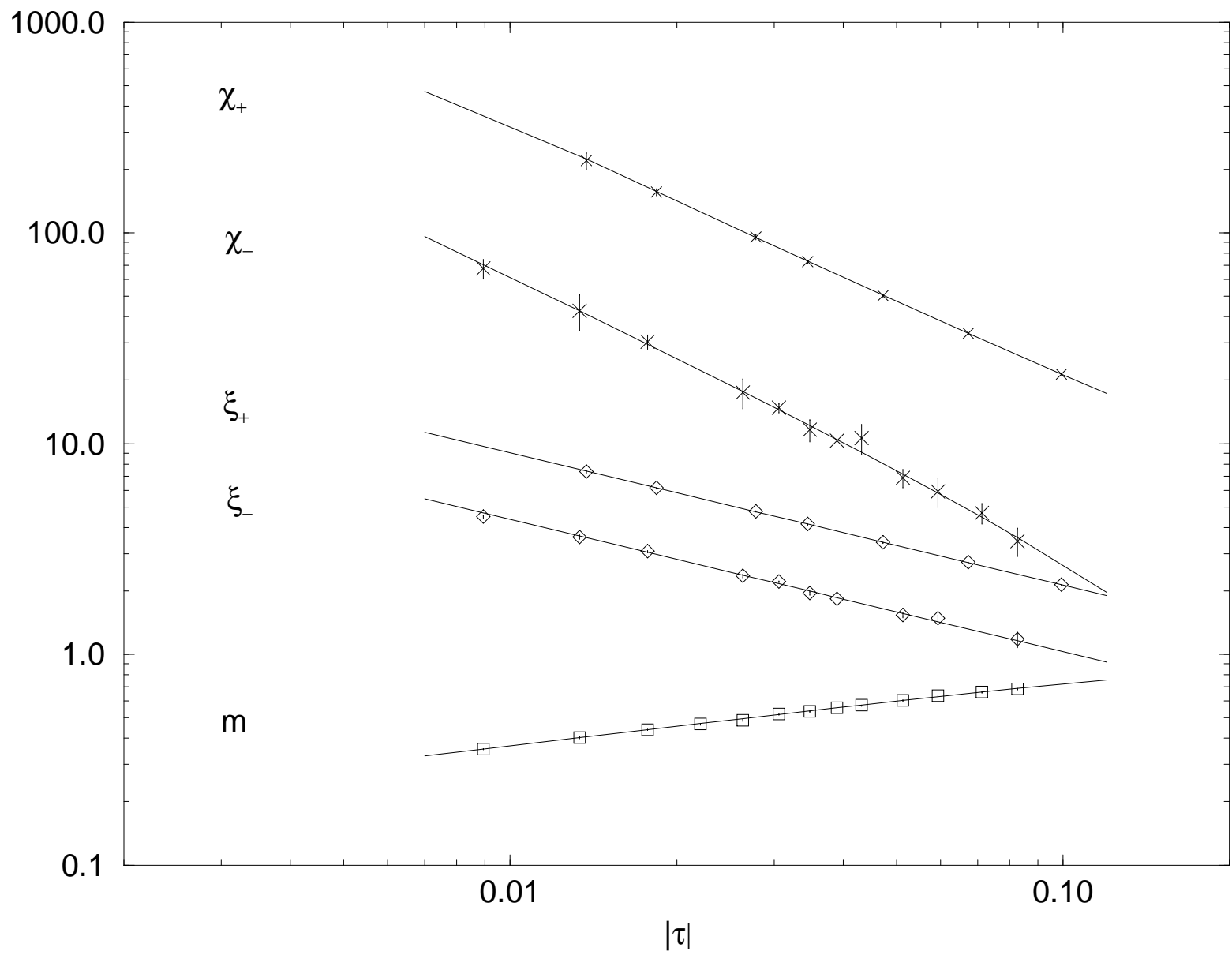
\end{document}